**Response of the Martian Ionosphere to Solar Activity including SEPs and ICMEs in a two week period starting on 25 February 2015**

F. Duru[1], D.A. Gurnett[1], D. D. Morgan[1], J. Halekas[1], R. A. Frahm[2], R. Lundin[3], W. Dejong[4], C. Ertl[4], A. Venable[4], C. Wilkinson[4], M. Fraenz[5], F. Nemec[6], J. E. P. Connerney[7], J. R. Espley[7], D. Larson[8], J. D. Winningham[2], J. Plaut[9] and P. R. Mahaffy[7]

(1) Dept. of Physics and Astronomy, University of Iowa, Iowa City, IA 52242, USA.

(2) Southwest Research Inst., PO Drawer 28510, San Antonio, TX 28228, USA.

(3) Swedish Institute of Space Physics, Box 812, SE-981 28, Kiruna, Sweden.

(4) Dept. of Physics, Coe College, Cedar Rapids, IA, 52402, USA.

(5) Max Planck Institute for Solar System Research, 37077 Gottingen, Germany.

(6) Faculty of Mathematics and Physics, Charles University in Prague, Czech Republic.

(7) NASA/Goddard Space Flight Center, Greenbelt, MD 20771, USA.

(8) University of California at Berkeley, Berkeley, CA, USA.

(9) Jet Propulsion Laboratory, M/S 183-601, 4800 Oak Grove Drive, Pasadena, CA 91109, USA.

Key points:

Solar activity in a two week period starting from 25 February 2015 altered the conditions in the Martian ionosphere.

The solar wind conditions and nightside ionosphere of Mars are studied simultaneously, with MAVEN, MEX and Mars Odyssey.

High ion escape rates and enhanced densities are observed during SEP enhancements and ICME.

**Abstract**

In a two-week period between February and March of 2015, a series of interplanetary coronal mass ejections (ICMEs) and (solar energetic particles) SEPs made contact with Mars. The interactions were observed by several spacecraft, including Mars Express (MEX), Mars Atmosphere and Volatile Evolution Mission (MAVEN), and Mars Odyssey (MO). The ICME disturbances were characterized by an increase in ion speed, plasma temperature, magnetic field magnitude, and energetic electron flux. Furthermore, increased solar wind density and speeds, unusually high local electron densities and high flow velocities were detected on the nightside at high altitudes during the March $8^{th}$ event. These effects are thought to be due to the transport of ionospheric plasma away from Mars. The peak electron density at periapsis shows a substantial increase, reaching number densities about $2.7 \times 10^4\,cm^{-3}$ during the second ICME in the deep nightside, which corresponds to an increase in the MOHigh-Energy Neutron Detector flux, suggesting an increase in the ionization of the neutral atmosphere due to the high intensity of charged particles. SEPs show a substantial enhancement before the shock of fourth ICME causing impact ionization and absorption of the surface echo intensity which drops to the noise levels. Moreover, the peak ionospheric density exhibits a discrete enhancement over a period of about 30 hrs around the same location, which maybe due to impact ionization. Ion escape rates at this time are calculated to be in the order of $10^{25}$ - $10^{26}\,s^{-1}$, consistent with MAVEN results, but somewhat higher.

**Introduction**

In a two-week period between February 25, 2015 and March 13, 2015, a series of interplanetary coronal mass ejections (ICMEs) were detected by multiple spacecraft studying the ionosphere and solar wind environment of Mars. At least four ICMEs were observed during this period, the last of which, on March $8^{th}$, was very strong. Using the data from six instruments on three different spacecraft, we first investigate the two-week period as a whole, then we study the effects of the ICME events on the solar wind environment and ionosphere of Mars focusing on the final March 8 event.

In the period starting on February 25, 2015 and ending on March 13, 2015, which has been studied extensively by the Mars Atmosphere and Volatile Evolution Mission (MAVEN) [*Jakosky et al.*, 2015], the orbit of the Mars Express (MEX) spacecraft had its periapsis in the deep nightside of Mars allowing examination of the effects of an ICME in this region. During the same period, the MAVEN spacecraft scanned solar zenith angles (SZA) between about 38° and 142°, being on the dayside during the strongest event. Finally, the Mars Odyssey (MO) spacecraft periodically covered a region between about 63° and 117° SZA. This high level of coverage makes it possible to perform a comprehensive investigation of this series of ICME interactions , comparing the effects observed in different regions of Mars.

Studying space weather events provides important insight into the escape processes and the evolution of the planetary atmospheres. Coronal mass ejections (CMEs) are massive bursts of plasma and magnetic field originating from the Sun's corona [*Howard et al.*, 2011]. They are termed ICMEs as they propagate through the interplanetary medium. ICMEs are observed to propagate radially away from the Sun at speeds greater than the typical solar wind velocity and are often preceded by a shock wave. A strong CME and a strong flare are associated with the emission across a wide range of the electromagnetic and energetic particle spectra. It is known that the atom and ion escape processes are enhanced when ICMEs interact with the atmospheres and ionospheres of unmagnetized planets [*Dubinin et al.*, 2009; *Edberg et al.*, 2010].

The discovery and description of ICMEs have a long and complex history (see, e. g, the historical summary by Howard, 2014). The characteristics of ICMEs are usually described in terms of probabilities because all of the characteristics of an ICME are rarely observed all together. The root cause of a CME is the sudden release of magnetic energy bound up in the solar corona. After initiation in the corona, the CME develops into a rapidly moving mass of solar particles. Propagation speeds are usually 300-1000 km/s but can be higher, with the total mass of the ejected material typically $10^{11}$-$10^{12}$ kg [*Howard*, 2014]. This mass of particles moves at speeds significantly greater than the ambient solar wind and can develop a detectable shock front [*Forbes et al.*, 2006]. The ICME catalog described by *Chi et al.* [2015] shows virtually all ICMEs with velocity greater than 400 km/s developing a shock at Earth orbit. After the shock, there is often a region of disturbed plasma, called the sheath, followed by a traveling flux rope, also known as a magnetic cloud [*Howard,* 2014]. The magnetic cloud can be rarefied compared to the surrounding plasma. We say "often" because the flux rope characteristic is not always observed; furthermore, it is not clear whether they are not observed because of observational issues or because the characteristics are not there in all cases. Forbes et al. [2006] state that a lack of clearly observed flux rope morphology could be due to an innate difference between types of ICMEs, spacecraft trajectory through the body of an ICME, or complexity due to collisions among several magnetic structures. According to Chi et al. [2015], about one-third of observed ICMEs contain a structure meeting several criteria for a magnetic cloud, or traveling flux rope. Where there is a shock, ions are accelerated to solar energetic particle (SEP) energies (~1 MeV). Gopalswamy et al. [2002] make a strong case that ICMEs produce SEPs primarily when a slow ICME is overtaken by a faster one. The effect at a given planet is determined by the solar wind magnetic field connection between the interaction region and the planetary atmosphere. The specific effects of the intense magnetic fields and energetic particles at Mars are described by Opgenoorth et al. [2013] and Morgan et al. [2014]. These effects include energization of ionospheric electrons, intrusion of solar wind electrons into the ionosphere, compression and possible erosion of the ionosphere, and strong impulsive intensification of the magnetic field. Because solar flares sometimes occur in approximate coincidence with a CME at the Sun, impulsive SEP events can be associated with ICMEs, whereas shock-driven SEPs are observed as so-called gradual events [Forbes et al., 2006]. It is known that large SEPs are related to ICMEs [Reames, 1999]. When an ICME occurs the charged particles start to move along the magnetic field lines. Since SEPs move much faster than the ICME shock they arrive earlier than the shock if the magnetic field lines are connected to the planet. SEP events have been shown to cause enhanced electron density in the ionosphere, which is seen from attenuation of the surface reflection during space weather events (see, e. g., Morgan et al., [2006], Nemec et al., [2014]).

There have been several studies of space weather events at Mars. For example, *Crider et al.* [2005] concluded that there was a strong compression of the Martian ionosphere during the Halloween Superstorm of 2003. *Opgenoorth et al.* [2013] studied three space weather events on Mars concluding, again, that significant ionospheric compression and heavy ion energization occurs when ICMEs interact with the Mars ionosphere. More recently, the effects of a strong ICME at Mars were studied by *Morgan et al.* [2014] using MEX and MO. All these studies had to rely on extrapolations of the upstream solar wind conditions since no upstream solar wind monitor was available. Finally, *Jakosky et al.* [2015] showed the response of Mars to the ICME on March 8, 2015, using data which provided information about the upstream solar wind from MAVEN solar wind monitors. Data from the MAVEN Solar Wind Ion Analyzer (SWIA) [*Halekas et al.*, 2015] and the Magnetometer (MAG) [*Connerney et al.*, 2015] were used to generate the escape rates in the regions affected by the ICMEs. Their observations will be used for comparison purposes in this study.

**Instruments**

Data from six instruments on three spacecraft, MEX, MAVEN and MO, are studied in order to understand the effect of the ICMEs, which occurred from February 25 to March 13, day 56 to day 72, of 2015.

MEX has been in an orbit around Mars since December 25, 2005, with six instruments monitoring changes in the Martian environment. One of these instruments is the Mars Advanced Radar for Subsurface and Ionospheric Sounding (MARSIS) instrument, which is a low-frequency radar. MARSIS, which has a 40 m tip-to-tip dipole antenna, 7 m monopole antenna, a radio transmitter, receiver and digital processing system [*Picardi et al.,* 2004], provides ionospheric density profiles determined from the remote sounding [*Gurnett et al.,* 2005]. When an ionospheric profile is visible, the peak ionospheric density can be identified. MARSIS is also able to obtain the local density and the magnitude of the local magnetic field through local electron plasma frequency harmonics and electron cyclotron echoes observed on the ionograms, which are plots of echo intensity as a function of time delay and frequency [*Gurnett et al.,* 2005; *Duru et al.,* 2008; *Akalin et al.,* 2010]. In remote sounding, a short radio pulse with frequency f is emitted and the time delay of the returning echo is measured. This process is repeated for 160 quasi-logarithmic steps in frequencies between 100 kHz and 5.5 MHz. The waves that are incident normal to the reflective surface of the ionosphere are reflected back to the sounder. For normal incidence, the reflection occurs at the altitude where the frequency of the wave transmitted by the sounder is equal to the electron plasma frequency.

An example ionogram is shown in Figure 1, which is taken from *Duru et al.* [2010]. This plot shows the echo intensity as a function of time delay on the vertical axis, and frequency on the horizontal axis. At lower frequencies, between 1.0 and 1.85 MHz in this ionogram from November 11, 2007, an ionospheric echo is observed. The surface reflection is detected at greater frequencies. The highest frequency in the ionospheric echo, denoted by $f_p$(max) in the figure, is the peak electron density in the ionosphere.

The vertical, equally spaced lines in the upper left corners of the figure are harmonics of electron plasma oscillations, which are due to the excitation of electron plasma by the wave activity generated by the sounder [*Duru et al.,* 2008]. The spacing between two consecutive lines, which is measured using an electronic ruler with adjustable tick marks, provides the electron plasma frequency local to the spacecraft, $f_p$. The electron plasma frequency, in turn can be used to calculate the local electron density using $n_e = (f_p)^2/(8980)^2$, where $f_p$ is in Hz and $n_e$ is the local electron density in cm$^{-3}$. This is a very useful method, which provides the local electron density even when the fundamental of the local electron plasma frequency is below the lower limit, 0.1 MHz.

The equally spaced horizontal lines on the left side of the ionogram are electron cyclotron echoes. The time difference between these horizontal lines is used to calculate the magnitude of the local magnetic field [*Gurnett et al.*, 2008; *Akalin et al.*, 2010].

The MEX Analyzer of Space Plasmas and Energetic Atoms (ASPERA-3) instrument is a suite of plasma and neutral detectors, that consists of an electron spectrometer (ELS), an ion mass analyzer (IMA), a neutral particle detector (NPD), a neutral particle imager (NPI) and a digital processing unit [*Barabash et al.*, 2004; 2006]. ELS is an electron spherical top-hat analyzer with a 360°x4° field of view divided into 16 angular sectors and an energy range from about 1 eV to 20 keV measured in 127 logarithmic steps every 4 s [*Barabash et al.,* 2006; *Frahm et al.,* 2006]. IMA is an ion top-hat energy analyzer coupled with an elevation analyzer at its entrance and a magnetic momentum analyzer at its exit. The IMA field of view is 360° x ±45°, divided into 16 steps of elevation and 16 azimuthal sectors. IMA measures ion distributions from ~-25 eV/q to 20 keV/q in 96 energy steps, and mass per charge spectra up to about 40 amu/q [*Barabash et al.,* 2006]. The ion fluxes for different energies can be used to derive ion flow velocities, including the solar wind speed.

MAVEN, which arrived at Mars in September 2014, has the main goals of studying the solar wind interactions of the upper atmosphere, and investigating atmospheric evolution and escape processes [*Jakosky et al.,* 2015]. MAVEN consists of three science packages. Its particle and fields package consists of a Solar Wind Electron Analyzer (SWEA) [*Mitchell et al.*, 2016], a Solar Wind Ion Analyzer (SWIA) [*Halekas et al,* 2013], a Solar Energetic Particle instrument (**SEP**) [*Larson et al.,* 2015], Suprathermal and Thermal Ion Composition instrument (STATIC) [*McFadden et al.,* 2015], a Langmuir Probe (LPW) [*Andersson et al.*, 2015] and a Magnetometer (MAG) [*Connerney et al.,* 2015a; 2015b]. MAVEN's Imaging Ultraviolet Spectrometer (IUVS) is a remote sounding package and its Neutral Gas and Ion Mass Spectrometer (NGIMS) package is the mass spectrometry instrument. In this paper, we present data from SWIA, which provides ion flow measurements in the magnetosheath, upstream solar wind and magnetotail. The toroidal energy analyzer and elevation analyzer on SWIA provides a field of view of 360° x 90° on a 3-axis stabilized spacecraft [*Halekas et al.,* 2013]. MAG, which consists of two independent, tri-axial fluxgate magnetometer sensors, measures the magnetic field in the solar wind and ionosphere of Mars [*Connerney et al.,* 2015a]. Some of the issues about MAG, such as compensating for spacecraft magnetic fields and verifying the accuracy of the measurements for weak fields are explained in *Connerney et al.* [2015b]. **SEP** measures energetic electrons (between 25 keV to 1 MeV) and protons (between 25 keV and 6 MeV) with 4 telescopes using dual-side solid state crystal technology [*Larson et al.,* 2015].

Finally, the High-Energy Neutron Detector (HEND) is part of the Gamma Ray Spectrometer suite of instruments onboard the MO spacecraft. Its main purpose is to study energetic neutrons [*Boynton et al.,* 2004], however, it has a special channel designed to detect X-rays and charged particles. We use this channel as a solar energetic particle detector.

**Data**

The data from MEX and MAVEN during the disturbed period from day 56 (Feb. 25) of 2015 to day 72 (March 13) of 2015 are presented in Figure 2 as a time series in six stacked plots. Panel (a) displays the solar wind speed obtained from spectral fits to the ion mass analyzer data on ASPERA-3, which is able to provide the solar wind speed when the solar wind $H^+$ kinetic energy is above 1 keV or when there is a clean $He^{++}$ signal. The speed is almost constant at about 400 km/s at the beginning of the period. Towards the end of day 57, starting from 22:15 UT, a small increase is observed. The solar wind speed gradually goes back to about 400 km/s at the end of the day 62, at about 18:16 UT it abruptly increases by about 50%. It stays almost constant up to about day 68, except for a small peak at day 66. Around 15:20 UT on day 67 (March 8[th]), the solar wind speed suddenly increases to 1000 km/s, from which it rapidly decreases. Each rise in the solar wind speed, marked by a dashed line, is an indication of a shock associated with the ICMEs. The solar wind speeds obtained by SWIA display similar behavior for this period [*Jakosky et al.,* 2015].

Panel b of Figure 2 shows the solar wind density obtained with SWIA on MAVEN. The ICME impacts on the ionosphere are associated with increases in the solar wind plasma density. At the start of the interval, the spacecraft is in the undisturbed solar wind, where the plasma density is ~5 $cm^{-3}$. Near the first solar wind velocity increase detected by ASPERA-3 IMA, the plasma density reaches ~25 $cm^{-3}$. Near the second such increase, the peak is ~20 $cm^{-3}$. Between day 66 and 71, which includes the third and fourth event, the density has four distinct peaks with values changing between 8 $cm^{-3}$ and 4 $cm^{-3}$. The larger of these peaks are nearly coincident with the solar wind velocity increases detected by ASPERA-3 IMA shown in Panel (a).

The magnetic field magnitude from MAG is shown in Panel (c), which also exhibits peaks closely following the dashed lines. In order to make sure that the spacecraft is measuring the solar wind magnetic field, we used the bow shock model from *Vignes et al.* [2000], obtained using the Mars Global Surveyor (MGS) observations to filter out data not taken when MAVEN is in the solar wind. For every orbit a series of the magnetic field vectors are obtained and the smallest vector magnitude has been chosen to be displayed. All four ICME shock times correspond to a peak in the magnitude of the magnetic field. During the final event, a smaller peak is followed by a more substantial one reaching values of 16 nT.

In Panel (d) the solar energetic ion spectra for high energy ions, collected by the **SEP** instrument onboard MAVEN, are shown. For each ICME an enhancement in the ion flux is observed closely following each of the solar wind velocity increases. Note that the intensification in the highest energy ion flux is observed about a day before the shock of the strongest ICME, which is an expected behavior [*Jakosky et al.,* 2015]. Part of this enhancement can also be due to previous events. Also note, that the third solar wind velocity increase, which is very small, is associated with a very small enhancement in the energetic ion flux.

Panel (e) displays electron spectra for different energies from ASPERA-3 ELS. Again, the increase in the flux by about an order of magnitude can be observed at the times of the events.

Finally, the local electron density from MARSIS Active Ionospheric Sounding (AIS) is presented in Panel (f). Data are shown for altitudes above 1200 km in order to exclude the high densities in the ionosphere at low altitudes. Since there are multiple values for each orbit, the local electron densities for a given pass look like a vertical line on the time scale shown in Figure 2. Even with the lower limit on the altitude, high local electron densities are observed around the times of the ICMEs, especially for the last event. Also, during the last event, the magnetic field magnitude obtained with MARSIS through electron cyclotron echoes (not shown) reaches 40 nT, which is about 4 times the usual value, and it is about 55 nT on day 63 (March 4[th]) which corresponds to the second event.

The enhancement of SEPs, which penetrate deep into the ionosphere has been shown to  cause absorption of the ground reflection (see e.g. *Morgan et al.,* [2006], *Nemec et al.,* [2014]). The top panel of Figure 3 presents the surface echo intensity during the two week period in question (for more information on how surface reflection is calculated see *Nemec et al.* [2015]). The intensities are obtained from MARSIS remote sounding using altitudes between 350 and 450 km and SZA higher than 107°. The median intensity is taken for each orbit. The horizontal dashed line at about 4.5 x $10^{-14}$ V$^2$/m$^2$/Hz is the normally expected surface echo intensity without absorption. There is a large depletion between the start and end of the period, with two possible recoveries in between. The middle and bottom panels show the **SEP** electrons and ions, respectively. The enhancements in the **SEP** ions and electrons are in excellent agreement with the intensity reductions. The drops in the intensity especially at the times of intense high energy electrons is in accordance with the fact that the electrons penetrate deeper.

The behavior of the peak density in the ionosphere, obtained remotely by MARSIS sounder, and the HEND charged particle/X-rays intensities are shown in Figure 4. Panel (a) of this figure is the same as the top panel of Figure 2. It displays the solar wind speed obtained from ASPERA-3 for comparison purposes. The peak density at the periapsis from MARSIS remote ionospheric sounding is shown in the middle panel and indicates fluctuations. However, since the spacecraft is on the nightside, the density is always below $1.5 \times 10^4$ cm$^{-3}$ except one time at the end of day 63. At this time, which corresponds to the shock of the second ICME, the peak electron density becomes more than twice the highest peak density recorded during this time interval, reaching $2.7 \times 10^4$ cm$^{-3}$. This increase happens at the same time as the depletion in the surface intensity and enhancements in the SEP electrons and ions.

The bottom panel (c) displays the count rate of charged particles and X-rays from the HEND instrument, with energies between hundreds of keV up to several MeV. The count rate is just below $4 \times 10^3$ s$^{-1}$ for most of the period, and has a small peak for the 2$^{nd}$ ICME, which corresponds to the big peak in the peak electron density. This substantial increase in the peak ionospheric density can be explained by the increased charged particle fluxes in the solar wind, which increase the ionization. A strong peak is observed between days 65 and and 67, the period in which the 3$^{rd}$ ICME, that has the weakest effect on other instruments, impacted the ionosphere of Mars. However, this time range corresponds to the enhanced **SEP** electrons (see Figure 3).

A closer look at the MARSIS local electron density is provided in Figures 5 and 6. Each figure shows five orbits and the dashed line indicates the periapsis location on each of the passes, which is centered on all the orbits. The local electron density as a function of time of a selection of orbits between days 60 and 62 is displayed in Figure 5. Some orbits which are featureless and have a small amount of data are skipped. In addition, orbits on which the ionospheric sounder is not operating are missing. Though the electron density is highly variable (see *Gurnett et al.,* [2010]), in all five of these orbits the density shows an increasing trend as the spacecraft descends. According to a survey of the nightside of Mars through MARSIS local electron densities by *Duru et al.* [2011], the electron density values at the location of the spacecraft are generally very low, changing between 2500 cm$^{-3}$ (around the terminator region, in the altitude range between 200-400 km) and 40 cm$^{-3}$ (in the SZA range between 130-140° and altitude range between 1200-1400 km). The electron density values of this figure are within the error bars of the study mentioned above. The shock of the second ICME happens right after the pass of orbit 14,173. During the next two available orbits, the electron density is almost constant at ~100 cm$^{-3}$.

Figure 6 shows that the local electron densities at high altitudes around the time of the fourth ICME (the shock is seen at the end of March 8[th], day 67) are much higher than predicted. For the first orbit in the figure, the electron densities in the altitude range between 400 and 800 km reach values ~500 cm$^{-3}$, which is above the average for that range. The electron densities reach very high values on passes from orbits 14,186 and 14,188. The densities above 700 km reach values above 800 cm$^{-3}$, which is more than 5 times the average values noted in *Duru et al.* [2011]. Around this time the densities near periapsis, when available, are much lower than at higher altitudes.

To gain a more detailed understanding of the ICME interaction on the nightside, we focus on the last and strongest event which happened on March 8, 2015, and examine the ionograms from orbits around this time. They reveal that the electron plasma oscillations show unusually high spacings at the beginning of the pass, indicating high local electron densities at very high altitudes are observed. As the spacecraft descends, the local electron plasma density decreases. When the plasma oscillations (observed as vertical lines in the ionograms) get very weak or disappear, a very diffused ionospheric echo is observed. This diffused reflection, indicating a turbulent ionosphere, is present for several minutes. After a few minutes the diffused reflection becomes distinct and sharp.

An intriguing feature is present in the plots of peak electron density obtained through MARSIS remote sounding (marked by A, B and C in Figure 7). This "top hat" feature, which is present in three orbits at the time of the ICME, is observed when the ionospheric echo on the ionograms becomes extended in frequency resulting in high peak density values. This feature is observed in all three cases in the first three panels of Figure 6 and for all of these cases the crustal magnetic field is weak, with values less than 50 nT. The features do not correspond to the times of high local electron density (see the corresponding orbits in Figure 6). All of these top-hat features are closer to periapsis than high local electron density regions. The only correspondence is between times 07:02 and 07:04 UT in orbit 14,188, where high peak electron densities and high local electron densities are observed at the same time. However, this time is 5 minutes before the top-hat shaped feature. Further investigation (the bottom panel of Figure 7) shows this feature occurs at about the same SZA, local time (LT) and latitude for all three cases. The examination of the plots of echo intensity as a function of time and universal time at fixed frequency shows that the altitude of the peak electron density of the ionosphere is at about 160 km, which is the expected height of the ionosphere based on a Monte Carlo Model [*Lillis et al.,* 2009]. At the other times the height of the ionosphere is at about 140 km, which is about 20 km lower than usual, suggesting compression due to solar activity. In two of the cases, ASPERA-3 ELS detects electrons flowing away from the planet. This is consistent with the findings of a recent study by *Withers et al.* [2012], who state that peak altitudes of about 150 km are observed at SZA > 115° and much lower altitudes are detected during solar energetic particle events.

The "top-hat" features may indicate a cloud of plasma, which is being transported from the dayside plasma towards the nightside because of the strong solar wind, or these features may be a result of intense impact ionization due to very energetic particles. Cross-terminator enhancements in the peak density have been seen in the past; however, a top-hat like feature which lasts for several orbits is recorded for the first time.

Figures 8 and 9 provide data for two orbits on March 8$^{th}$, 2015, where the highest local electron densities are obtained. The bottom panels (Panels (d)) show the local electron densities determined from MARSIS. The flow velocities for $O^+$ (blue) and $O_2^+$ (green) ions calculated with ASPERA-3 data are given in Panels (c). Expected local electron density values for the given altitude and solar zenith angle range are indicated by the red horizontal lines, which are the average values taken from *Duru et al.* [2011]. The purple line marks the location of the top-hat feature for the given orbit. During the ICME, the local electron densities reach very high values, about eight times the expected density in some cases. The ion flow velocities are around $10 - 20$ km/s for the high density regions. Panels (a) and (b) show the total heavy ion flux and heavy ion outflow obtained from IMA, respectively.

**Escape Rates**

It is known that Mars has lost a significant amount of its atmosphere. A large fraction of this loss is believed to be due the escape of plasma. So far several studies have provided estimates of ion escape rates from the Martian ionosphere. According to *Barabash et al.* [2007], the highest rates are obtained for $O^+$ ions, reported to be $1.6 \times 10^{23}$ s$^{-1}$. A more recent study provides evidence of solar cycle effects on the ion escape rates stating that the average heavy ion escape rate is about $1 \times 10^{24}$ s$^{-1}$ during solar minimum and $1 \times 10^{25}$ s$^{-1}$ during solar maximum [*Lundin et al.,* 2013].

Here, we provide the heavy ion flux and corresponding heavy ion outflow from IMA shown on Panels (a) and (b) of Figures 8 and 9 as a lower limit for the escape rate. The ion outflow rate is about $8 \times 10^{23}$ ions/s$^{-1}$ at the beginning of pass 14,188. It increases after a drop. At around 7:13 UT on day 67 it reaches values as $1 \times 10^{24}$. At 20:50 UT (see Figure 9), which is after the shock of the fourth ICME, the value is as high as $3 \times 10^{24}$ ions/s. After this time, the outflow rate decreases to values around $10^{23}$ ions/s.

*Jakosky et al.* [2015] provided estimates of the ion escape rates using MAVEN data and a model during the March 8[th], 2015 ICME. We try to reproduce their results using only the data we have and making some assumptions. Knowing the local electron density, from electron plasma oscillations, and the ionospheric flow velocity, from ASPERA-3 IMA, we can make a rough estimate of the ion escape rates during the ICME event. Since MEX is in the deep nightside, it also gives us an idea about how the ion escape rates change on the nightside with the solar wind interaction occurring on the dayside. In order to calculate the ion escape rates, we assume that the escape from the ionosphere of Mars is cylindrically symmetric. Consider a thin disk around the planet starting at the location where we have the data (see Figure 10). The escape rate can be calculated by multiplying the average local density by the flow velocity and the area of the disk (given by $2\pi\rho\Delta\rho$, where $\rho$ is the radius of the thin shell and $\Delta\rho$ is its thickness). Ion escape rates are calculated for the high density regions for the two passes shown in Figures 8 and 9. For the first time period, which gives us the highest rates, the ion escape rate is found to be $9.1 \times 10^{25}$ s$^{-1}$ (using $\rho$=3371 km, $\Delta\rho = 700$ km and local $n_e$= 600 cm$^{-3}$) and for the second time period the ion escape rate is $3.2 \times 10^{25}$ s$^{-1}$ (using $\rho$=3371 km, $\Delta\rho = 400$ km and local $n_e$= 200 cm$^{-3}$). Both values are of the same order of magnitude as the results from *Jakosky et al.* [2015], who obtain ion escape rates $1.46 \times 10^{24}$ s$^{-1}$, $1.06 \times 10^{25}$ s$^{-1}$ and $3.34 \times 10^{25}$ s$^{-1}$ for three different stages of the ICME using MAVEN data and a model. The period for which we have determined the escape rates corresponds to the times just before the ICME shock and just after it. Our first value, $9.1 \times 10^{25}$ s$^{-1}$, obtained at about the same time *Jakosky et al.* [2015] computed the second value, $1.06 \times 10^{25}$ s$^{-1}$, and our second value, $3.2 \times 10^{25}$ s$^{-1}$, corresponds to the third case in *Jakosky et al.* [2015], $3.35 \times 10^{25}$ s$^{-1}$. Our first value is higher, but of the same order of magnitude as the MAVEN results, whereas our second time gives almost the exact same number for the escape rate. The values are about an order of magnitude higher than the numbers obtained by *Lundin et al.* [2013] for solar maximum times, which is expected during a strong ICME. The local electron densities at the given altitude and SZA range drop to the expected values (around 100 cm$^{-3}$). Since we do not have flow velocity data for the times outside the ICME, it is not possible to give exact numbers, but it is safe to assume that the escape rates should be about an order of magnitude less than the ICME times. The method used here provides an approximation of the escape rate determined making use of several assumptions. The fact that the flux of particles is not actually cylindrically symmetric and that not all the ions are above the escape velocity of Mars leave the planet makes our results only an approximation. Similar methods have been used previously; however, they yield results in the same order of magnitude ($10^{25}$ ions/s) for trans-terminator flow at Mars [*Fraenz et al.,* 2010].

**Discussion and Summary**

We have presented here a comprehensive study of a highly active and variable two-week period in the Martian ionosphere in 2015 using measurements from three different spacecraft. At least 4 ICMEs are detected during this time frame and their effects are observed by all three spacecraft in different regions around the planet: MEX being in the deep nightside, and MAVEN and MO on the dayside. The strongest of these ICMEs occurred on March 8[th]. Besides being the strongest ICME, it occurs during a more chaotic period of the ionosphere than the previous ICMEs.The effects of the previous ICMEs caused turbulent ionospheric conditions which persisted when the March 8[th] ICME hit Mars. Also, the SEP electrons and ions are enhanced before the shock of this last ICME. *Gopalswamy et al*. [2002] conclude that the shock of a CME accelerates SEPs from the solar wind affected by preceding CMEs, instead of from the quiet solar wind. This statement is consistent with the appearance of SEPs before a ICME, which is preceded by three others in this two week period.

As expected, all ICMEs are evident on examination of the solar wind speed obtained with ASPERA-3, which showed an increase in solar wind speed at each event indicating a shock. The solar wind speed increases are also consistent with peaks in the solar wind density and solar energetic particle flux from MAVEN.

The charged particle/X-ray count rate from HEND shows a narrow peak right after the second ICME event, and a very wide peak between the third and fourth ICME events, where the count reached values around $1 \times 10^4$ s$^{-1}$. This behavior is consistent with the high energy ion and electron flux observed in the **SEP** instrument data. Peak electron density values around the periapsis for each available pass are studied at the times when HEND shows count peaks. The peak electron density obtained from MARSIS remote sounding reveals that there was a substantial increase at the time of the narrow peak in the HEND data, which is believed to be due to an increase in the ionization of the neutral atmosphere due to increased fluxes of X-rays and charged particles.

The SEP may originate from a new original flare, independent from the ICME, but they may be observed before the ICME if the ICME traveling from the Sun is magnetically connected to Mars. *Jakosky et al.* [2015] suggest that the SEP enhancements observed before the ICME on March 8, are due to the coming ICME. As SEP particles arrive at Mars, they will penetrate deep into the ionosphere causing ionization and possible absorption. The study of the surface reflection echo intensity shows that this is what is actually happening during this two weeks. The intensity is highly depleted during this time range, which drops from values about $4.5 \times 10^{-14}$ $V^2/m^2/Hz$ to the noise level right around February 28 (day 59). Two possible recovery regions are observed peaking around day 62 and 65, which correspond to the low intensity regions on **SEP** ions and electron plots. As the SEPs are enhanced the surface echo intensity is attenuated to noise level.

The local electron density from MARSIS for the altitudes above 1200 km displays very high values exceeding 800 $cm^{-3}$ around the time of the fourth ICME. High flow velocities at the same time suggest transport of plasma from the dayside to the nightside of Mars because of the pressure of the high speed and dense solar wind. The highest densities correspond to the times before the shock of the ICME where the SEPs are enhanced. Although, impact ionization due to high energy solar wind particles remains as another possible reason for very high electron densities, it is very unlikely since the altitudes are very high and the plasma is tenuous.

Investigation of three orbits around the last event shows that in general the altitude of the ionosphere is about 140 km, which is about 20 km less than the predicted value for the nightside, suggesting a compression of the ionosphere. Other previous studies have also found a compressed ionosphere [*Crider et al.,* 2005; *Opgenoorth et al.,* 2013; *Ulusen et al.,* 2012, *Morgan et al.* 2014]. Recently, *Dong et al.* 2015, who performed analysis of solar wind interaction of Mars during the ICME on March 8th, reported a decrease in the altitude of both the bow shock and the magnetic pile-up boundary. In this last ICME event, an exception to this idea is found in the observation of a persistent feature seen in the ionospheric peak electron density for three orbits at SZA between 138° and 141° and local time of 21.3 hours. During these observations the altitude of the nightside ionosphere is about 160 km (the top-hat feature presented in Figure 6). No strong crustal fields are present in this region. This feature lasts about 2 minutes, which corresponds to a distance of about 500 km in the direction of the spacecraft, in each pass and is characterized by a hat shaped appearance on the peak electron density plots. The features observed on orbits 14,186 and 14,188 are before the shock of the ICME. However, they happen at the time of the very strong SEP ions and electrons signature. During one of the three times investigated outward flowing electrons are observed by ELS. The ionospheric density peaks for the time of the top-hat feature, which can be due to a localized and transient plasma cloud formed as a result of transport or impact ionization. Knowing that the SEPs (especially electrons) can penetrate deep in the ionosphere suggests that the density increase due to impact ionization is a likely scenario.

SEP enhancements are decreased considerably after the shock of the ICME, which can be explained with the fact that shocks alter the magnetic field topologies.

The IMA ion outflow rates are also reported to be used as a "lower limit" for the ion escape rates. It should be noted however that the field of view of IMA is limited and the energies below 50 eV are not scanned (the elevation analyzer voltage is set to zero for ions detected below 50 eV), which may lead to an overestimation of the velocities. Also, the lack of knowledge about the spacecraft potential can be misleading even though a potential correction is applied [*Fraenz et al.*, 2015]. The ion flux from IMA changes in the range between $2 \times 10^5$ and $2 \times 10^7$ $cm^{-2}s^{-1}$ on day March 8th. The ion outflow for the two times yielding highest local electron densities from MARSIS are in the order of $10^{23}$-$10^{24}$ ions/s.

With the aim of confirming the results reported in Jakosky et al., 2015 for the time of the ICME on March 8th, we calculated the ion escape rates using the local electron density and flow velocity. The data from two orbits, one just before the ICME and other after the shock are used. For the first time period the ion escape rate was found to be 9.1 x $10^{25}$ $s^{-1}$ and for the second time period it is 3.2 x $10^{25}$ $s^{-1}$. The first escape rate we obtain is less than one order of magnitude higher than the one calculated by *Jakosky et al.* [2015] for the same period. It should be kept in mind that the technique used by them does not take into account the SEPs, which we believe is what increased the local electron density values even before the shock of the ICME. The second value is essentially the same as that of *Jakosky et al.* [2015], suggesting that our method with its assumptions works reasonably well.

According to *Lundin et al.* [2013], the average heavy ion escape rates during solar minimum and maximum are $1\times10^{24}$ and $1\times10^{25}\,s^{-1}$, respectively. During solar storm times it is expected to have one or two order of magnitude higher escape rates. Even though the method used is approximate, the results are reasonable when compared with the average ion escape rates, keeping in mind that the ion escape rates that were generated by the March $8^{th}$ ICME are instantaneous values measured at a single point during a highly dynamic period.

**Acknowledgement**

This paper describes observations obtained from both the NASA sponsored MAVEN and Mars Odyssey, and the ESA sponsored Mars Express mission to Mars. Data from Mars Express AIS and ASPERA-3 ELS and IMA are available from the NASA Planetary Data System (PDS) Geosciences Node (http://pds-geosciences.wustl.edu/). Data from Mars Odyssey Gamma Ray Spectrometer, High Energy Neutron Detector and MAVEN are also available from the NASA PDS Geosciences Node. We are grateful to these agencies for making this study possible. This work is funded by contract 1224107 administered by the Jet Propulsion Laboratory at the University of Iowa, NASA contract NASW-00003 at Southwest Research Institute, and the Swedish National Space Board.

**Figures Captions:**

**Figure 1:** Sample ionogram from November 11, 2007 (from Duru et al. 2010, Figure 1b). The received intensity is indicated by the plotted color according to the color bar. The electron plasma oscillations, electron cyclotron echoes, ionospheric echo, surface reflection, and peak density in the ionosphere ($f_p$(max)) are shown.

**Figure 2:** Time series of several measured quantities during the two-week period from Feb. 25 to March 13, 2015. The observations indicate that four ICME impacts at Mars, marked by the dashed lines, occurred during this period. Panel (a): Solar wind speed from ASPERA-3 IMA. Panel (b): Solar wind density from SWIA. Panel (c): Magnitude of magnetic field in the solar wind from MAG. Panel (d): High energy ion flux from **SEP**. Panel (e): Electron flux from ASPERA-3 ELS. Panel (f): Local electron density above 1200 km from MARSIS. [Panels (b) and (d) are taken from Jakosky et al., 2015.]

**Figure 3:** Top panel: The surface echo intensity in the altitude range between 350 and 450 km and SZA > 107° during the two week period in question. The dashed-line indicates expected

surface echo intensity when there is no absorption. Middle panel: High energy electron flux from **SEP**. Bottom panel: High energy ion flux from **SEP.**

**Figure 4:** The same two week period as in Figure 2, which displays 4 ICME events at Mars indicated by the dashed lines. Panel (a) : Solar wind speeds obtained from ASPERA-3 IMA (kept for reference purposes). Panel (b): Peak electron density around periapsis from MARSIS remote sounding. Panel (c): Charged particle/X-ray count from HEND.

**Figure 5:** The local electron plasma frequency as a function of universal time (UT) and altitude are shown for five passes: 14,165 (day 60), 14,168 (day 61), 14,170 (day 62), 14,171 (day 62) and 14,173 (day 62). Periapsis is indicated by the dashed line.

**Figure 6:** The local electron plasma frequency as a function of universal time (UT) and altitude are shown for five passes: 14,184 (day 66), 14,186 (day 66), 14,188 (day 67), 14,190 (day 67) and 14,191 (day 68). Periapsis is indicated by the dashed line.

**Figure 7:** Panel (a), (b) and (c) show peak electron density in the ionosphere from MARSIS remote sounding for orbits 14,186, 14,888 and 14,190, respectively. Panel (d) presents the SZA and LT of the hat-shaped features present in three of the orbits. The beginning and end of the feature is marked and joined by a line for each pass.

**Figure 8:** Panel (a): Total heavy ion flux from ASPERA-3 IMA for the orbit 14,188. Panel (b): Heavy ion outflow from IMA Panel (c): O+ (blue) and O2+ (green) flow velocities obtained from IMA. Panel (d): Corresponding local electron densities from MARSIS. The red line indicates the expected local electron density values in the given SZA and altitude range obtained from Duru et al. [2011]. The thick purple lines on the local electron density plots indicate the location of the top hat feature in these orbits.

**Figure 9:** Panel (a): Total heavy ion flux from ASPERA-3 IMA for the orbit 14,190. Panel (b): Heavy ion outflow from IMA Panel (c): O+ (blue) and O2+ (green) flow velocities obtained from IMA. Panel (d): Corresponding local electron densities from MARSIS. The red line indicates the expected local electron density values in the given SZA and altitude range obtained from Duru et al. [2011]. The thick purple lines on the local electron density plots indicate the location of the top hat feature in these orbits.

**Figure 10:** The sketch of the ionosphere and solar wind environment of Mars. A thin disk is defined around the Mars to calculate the escape rate of ions.

**Figures:**

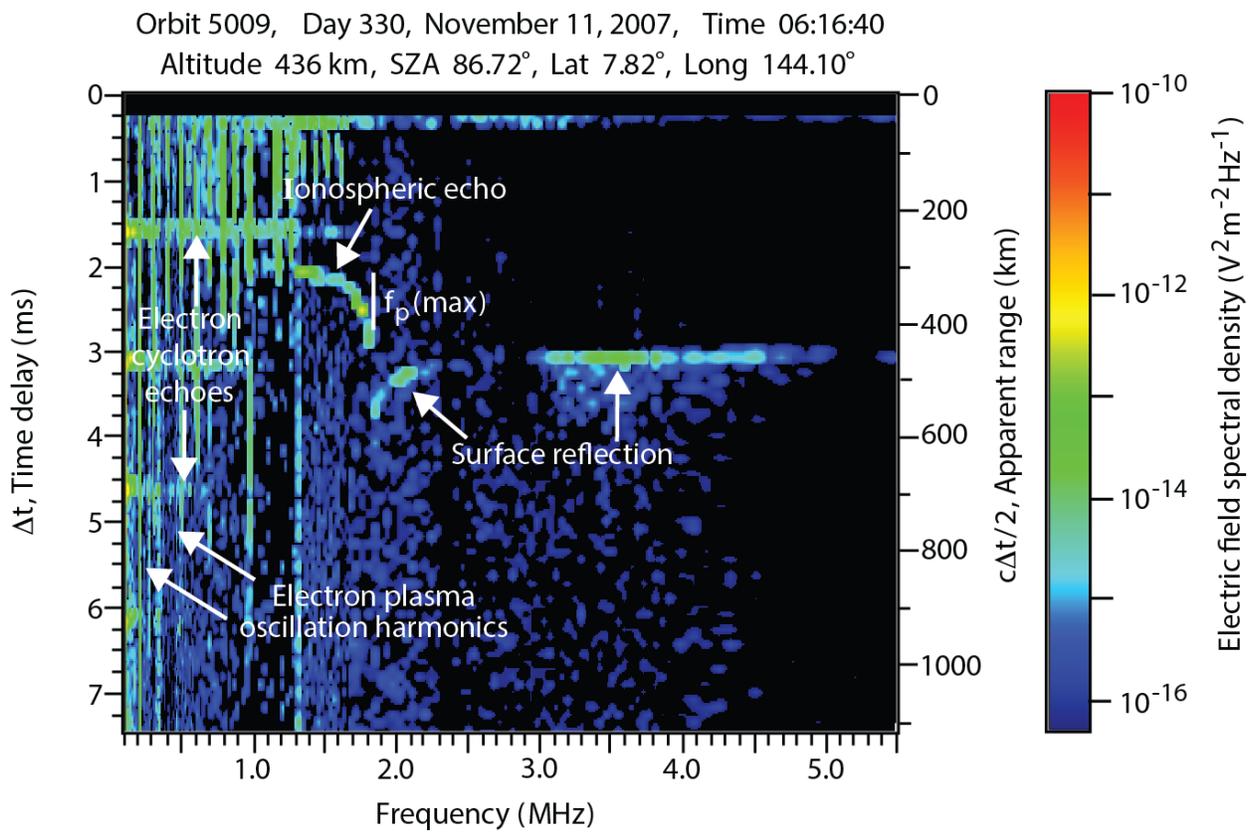

Orbit 5009,   Day 330,  November 11, 2007,   Time  06:16:40
Altitude  436 km,  SZA  86.72°,  Lat  7.82°,  Long  144.10°

**Figure 1**

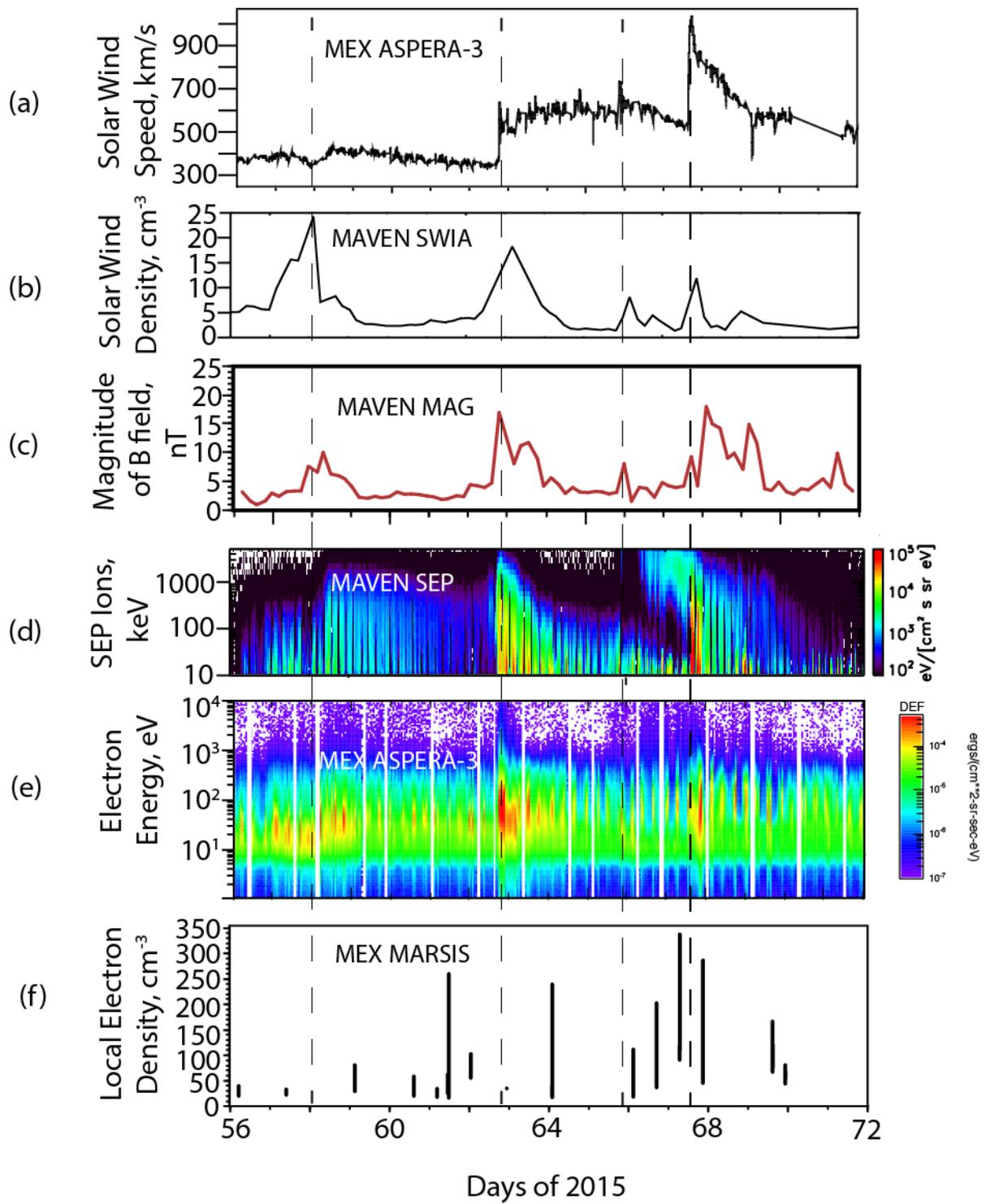

**Figure 2**

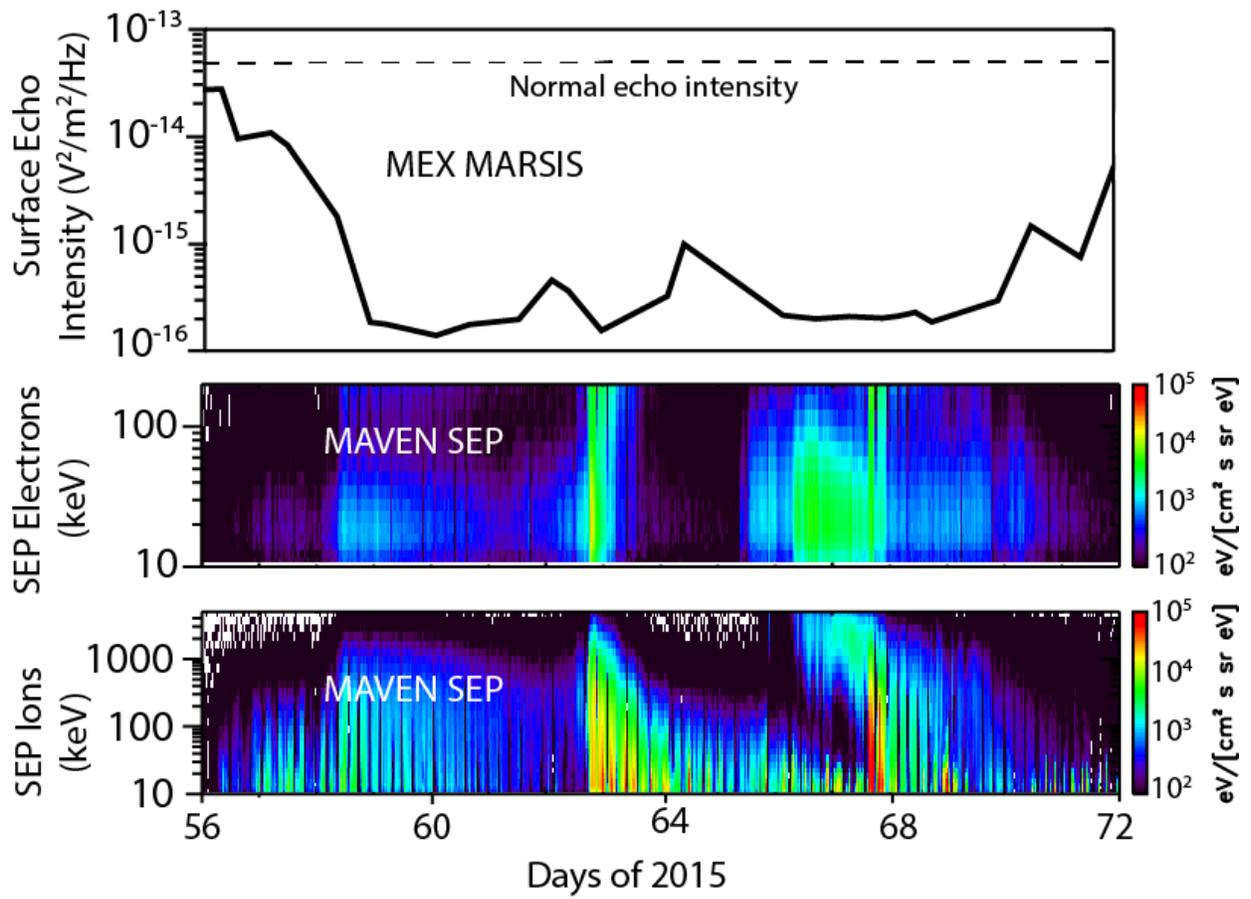

**Figure 3**

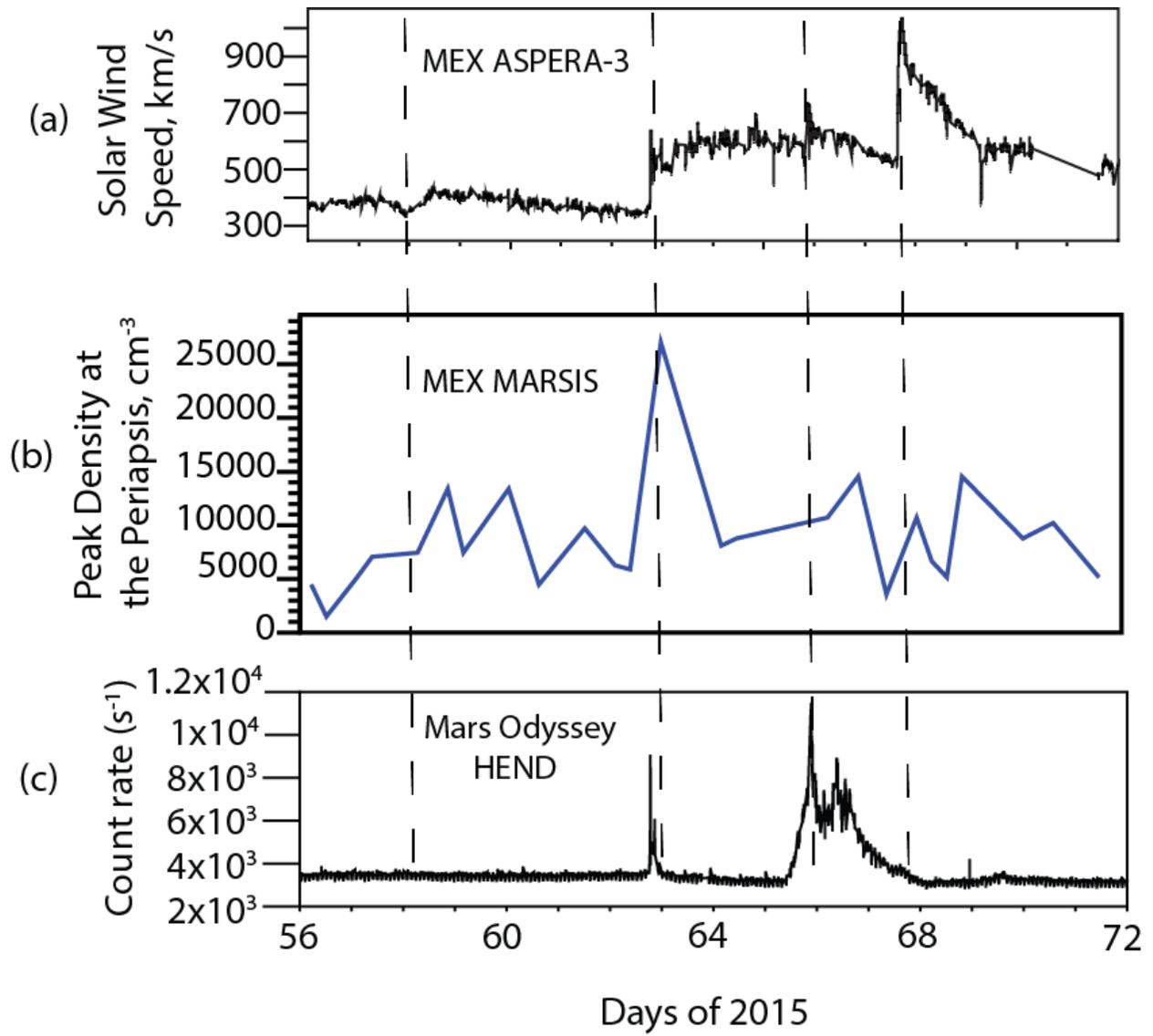

**Figure 4**

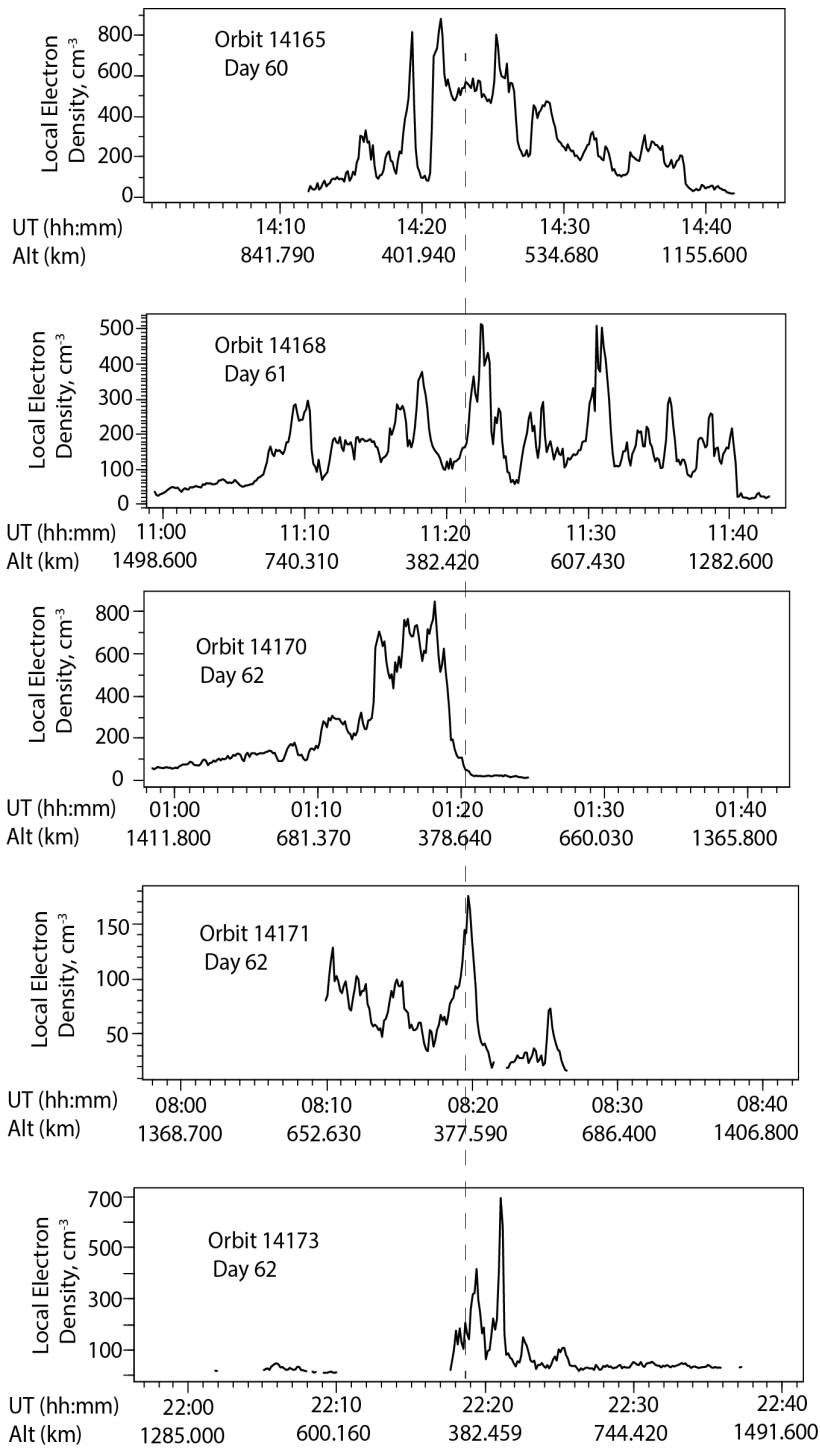

**Figure 5**

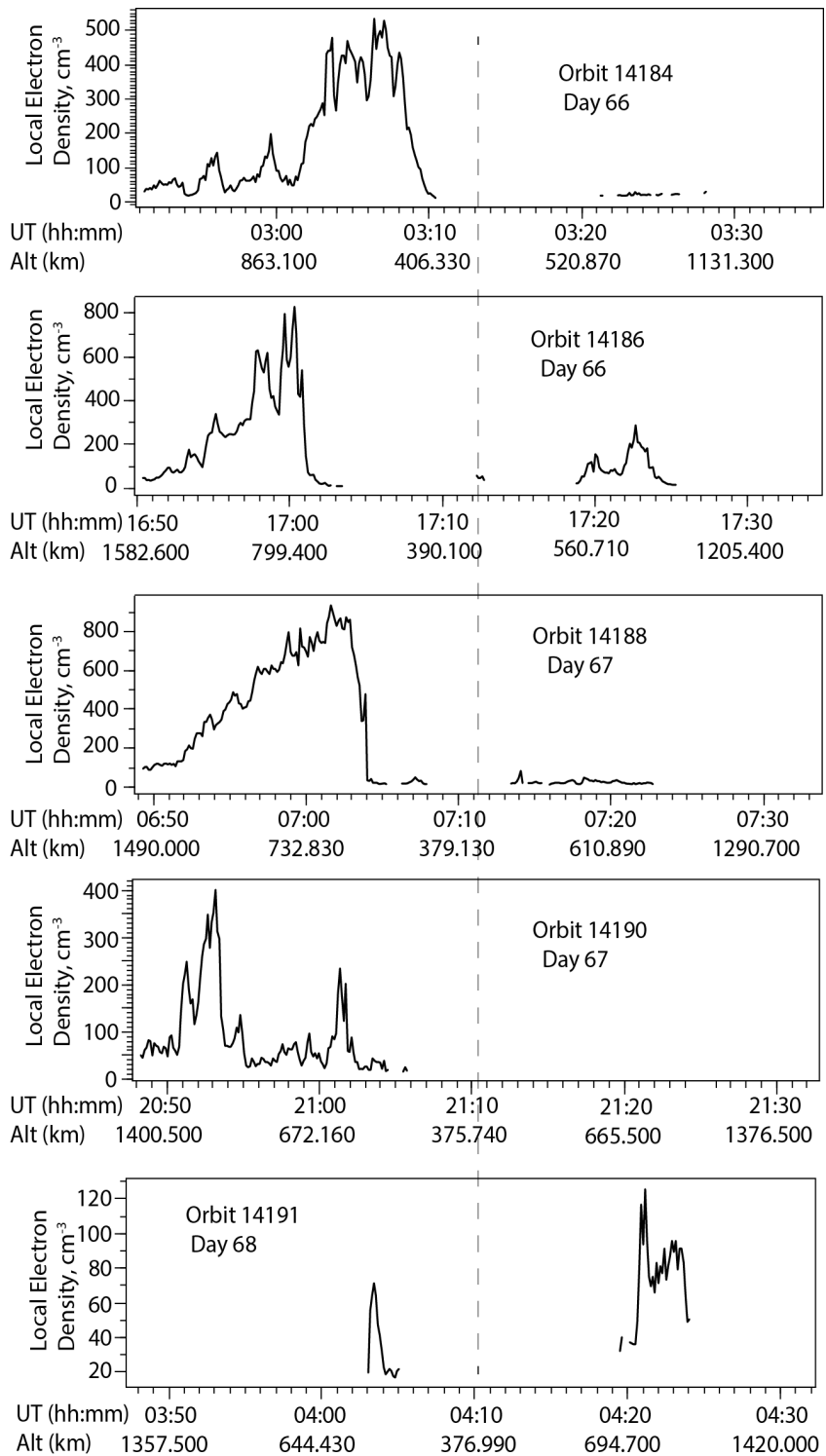

**Figure 6**

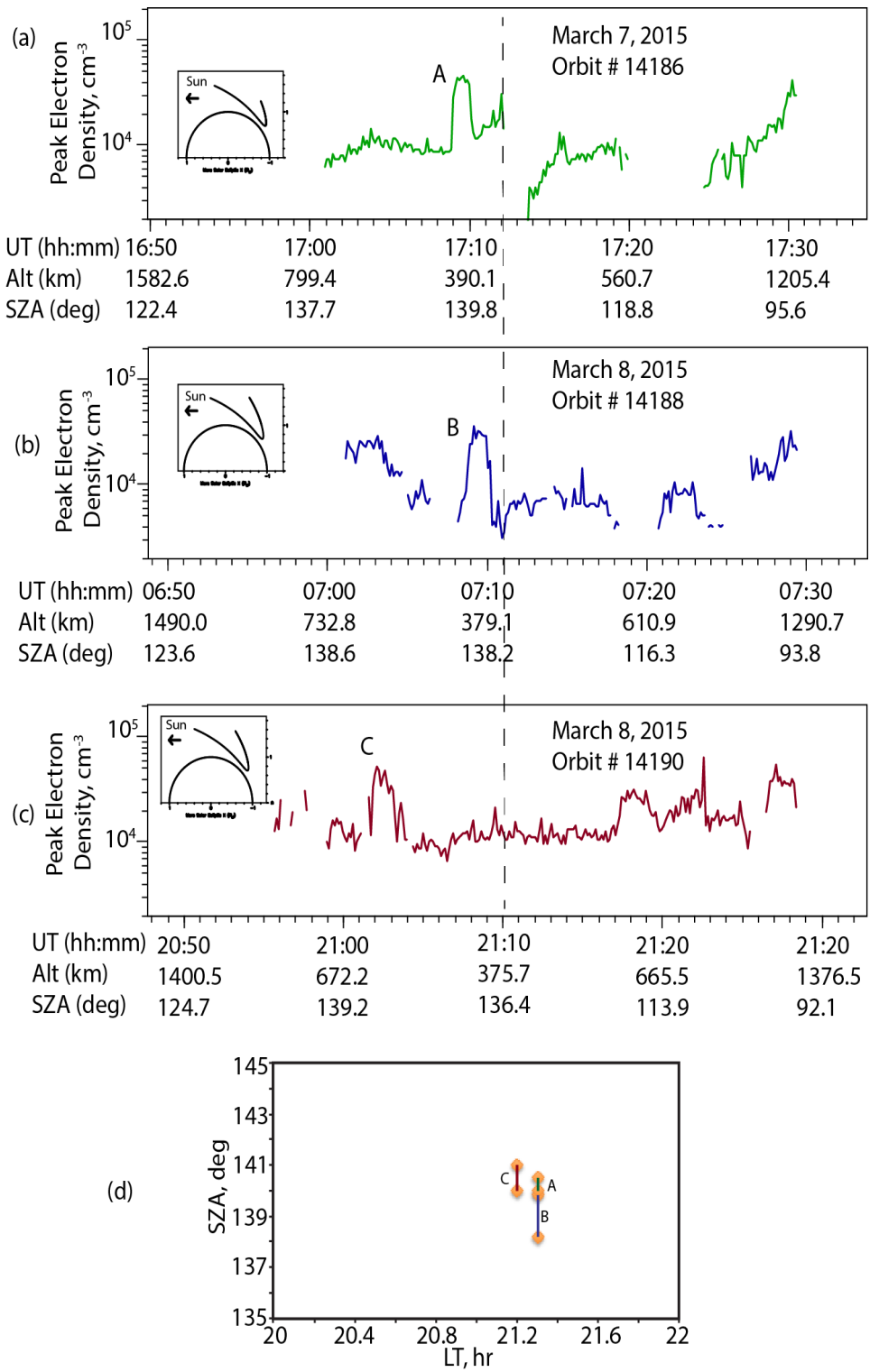

**Figure 7**

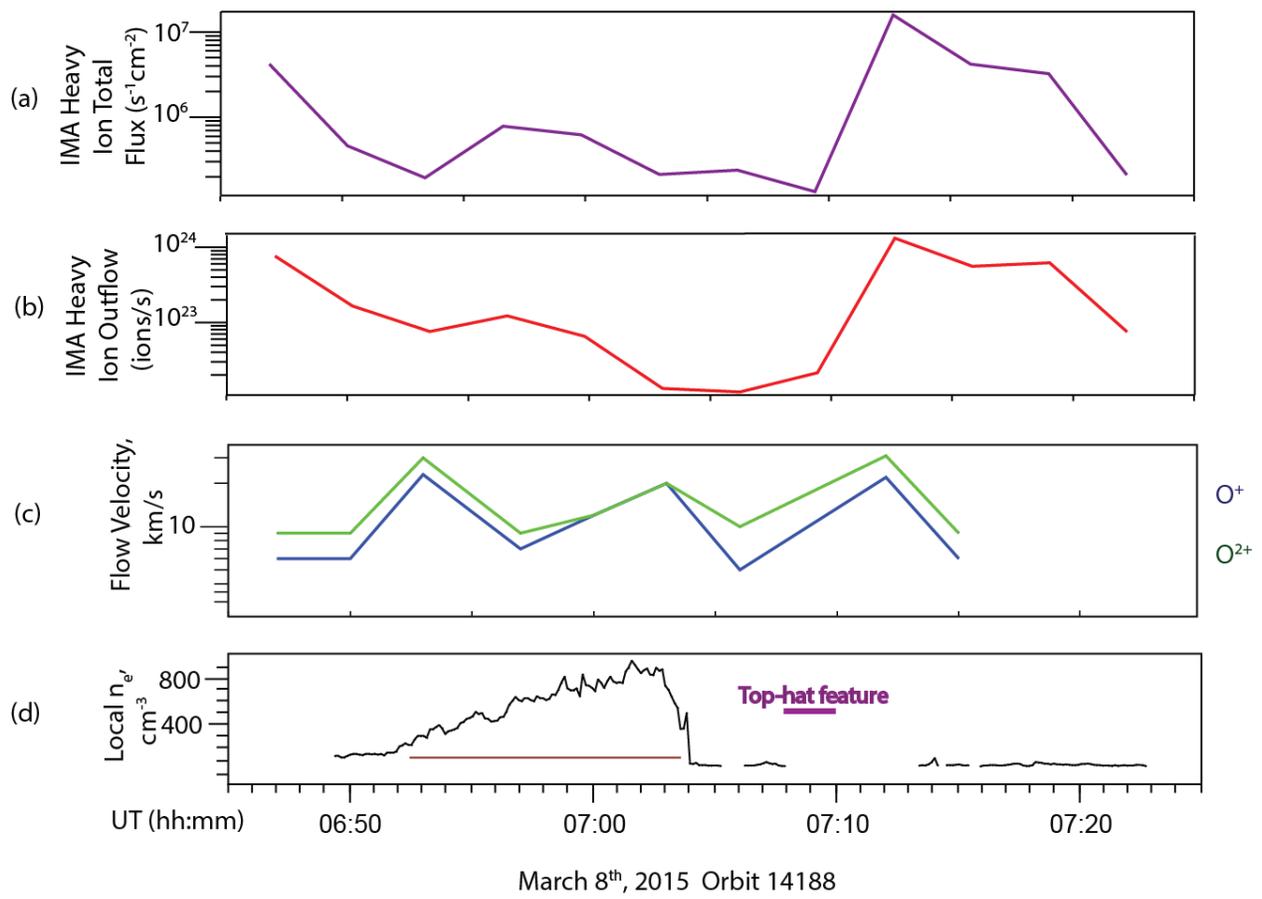

**Figure 8**

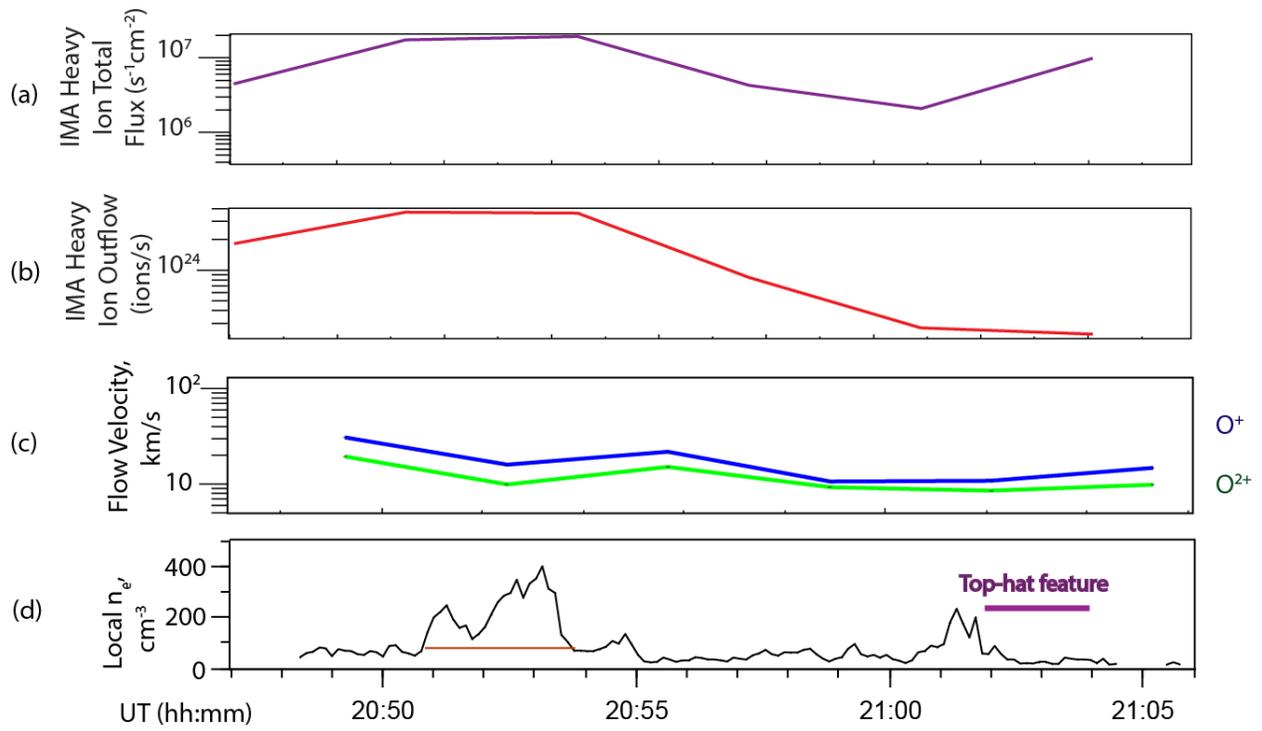

**Figure 9**

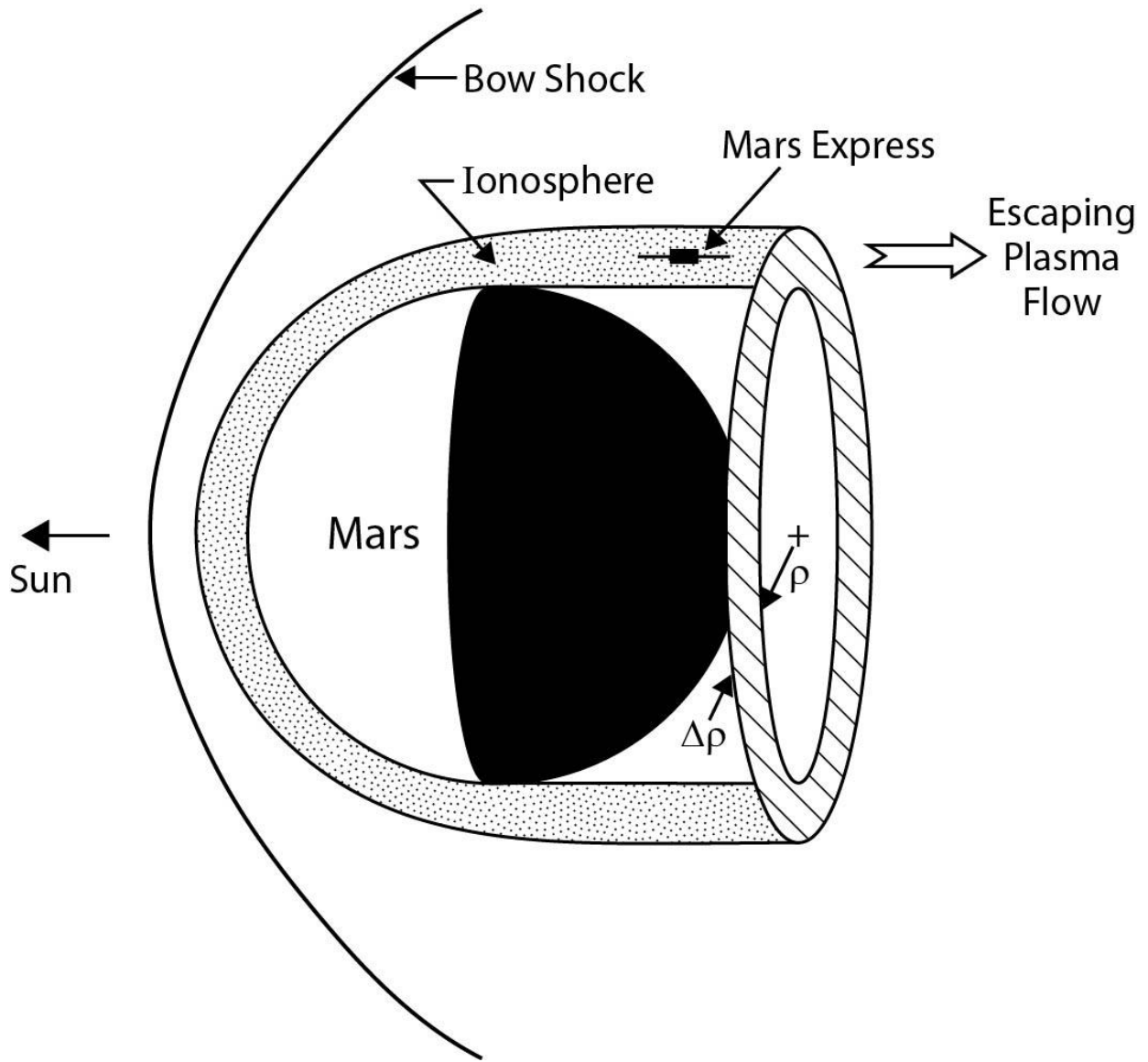

**Figure 10**